\RequirePackage[l2tabu, orthodox]{nag}

\documentclass[a4paper,UKenglish]{lipics}

\usepackage{fixltx2e}[2006/09/13]

\usepackage[utf8]{inputenc}

\usepackage{amsmath}
\usepackage{amsfonts}
\usepackage{graphicx}

\usepackage[vlined, ruled, noend, linesnumbered]{algorithm2e}

\newcommand{\C}{\mathcal{C}}
\newcommand{\T}{\mathcal{T}}
\newcommand{\wt}{\widetilde}
\newcommand{\troot}{\mathrm{root}}
\newcommand{\children}{\mathrm{children}}
\newcommand{\softO}{\widetilde{O}}
\newcommand{\hide}[1]{\relax}
\renewcommand{\epsilon}{\varepsilon}

\title{Multiple-Edge-Fault-Tolerant Approximate Shortest-Path Trees\footnote{This work was partially supported by the Research Grant PRIN 2010 ``ARS TechnoMedia'', funded by the Italian Ministry of Education, University, and Research.}}
\titlerunning{Multiple-Edge-Fault-Tolerant Approximate Shortest-Path Trees}

\author[1]{Davide~Bil\`o}
\author[2]{Luciano~Gual\`a}
\author[3,4]{Stefano~Leucci}
\author[3,5]{Guido~Proietti}
\authorrunning{D. Bil\`o, L. Gual\`a, S. Leucci, and G. Proietti}

\affil[1]{DUMAS, Università di Sassari, Italy. \texttt{davide.bilo@uniss.it}}
\affil[2]{DII, Università di Roma ``Tor Vergata'', Italy. \texttt{guala@mat.uniroma2.it}}
\affil[3]{DISIM, Università degli Studi dell'Aquila, Italy. \texttt{\{name.surname\}@univaq.it}}
\affil[4]{DI, ``Sapienza'' Università di Roma, Italy.}
\affil[5]{IASI, CNR, Roma, Italy.}

\Copyright{Davide Bil\`o, Luciano Gual\`a, Stefano Leucci, and Guido Proietti}

\serieslogo{}
\volumeinfo
  {}
  {0}
  {}
  {1}
  {1}
  {1}
\EventShortName{}
\DOI{}

\begin{document}

\maketitle	
\begin{abstract}
Let $G$ be an $n$-node and $m$-edge positively real-weighted undirected graph. For any given integer $f \ge 1$, we study the problem of designing a sparse \emph{f-edge-fault-tolerant} ($f$-EFT) $\sigma${\em -approximate single-source shortest-path tree} ($\sigma$-ASPT), namely
a subgraph of $G$ having as few edges as possible and which, following the failure of a set $F$ of at most $f$ edges in $G$, contains paths from a fixed source that are stretched at most by a factor of $\sigma$.
To this respect, we provide an algorithm that efficiently computes
an $f$-EFT $(2|F|+1)$-ASPT of size $O(f n)$. Our structure improves on a previous related construction designed for \emph{unweighted} graphs, having the same size but guaranteeing a larger stretch factor of $3(f+1)$, plus an additive term of $(f+1) \log n$.

Then, we show how to convert our structure into an efficient $f$-EFT \emph{single-source distance oracle} (SSDO), that can be built in $\softO(f m)$ time, 
has size $O(fn \log^2 n)$, and is able to report, after the failure of the edge set $F$, in $O(|F|^2 \log^2 n)$ time a $(2|F|+1)$-approximate distance from the source to any node, and a corresponding approximate path in the same amount of time plus the path's size. Such an oracle is obtained by handling another fundamental problem, namely that of updating a \emph{minimum spanning forest} (MSF) of $G$ after that a \emph{batch} of $k$ simultaneous edge modifications (i.e., edge insertions, deletions and weight changes) is performed. For this problem, we build in $O(m \log^3 n)$ time a \emph{sensitivity} oracle of size $O(m \log^2 n)$, that reports in $O(k^2 \log^2 n)$ time the (at most $2k$) edges either exiting from or entering into the MSF.
As a result of independent interest, it is worth noticing that our MSF oracle can be employed to handle arbitrary sequences of $o(\sqrt[4]{n}/\log n)$ (non-simultaneous) updates with a worst-case time per update of $o(\sqrt{n})$. Thus, for relatively short sequences of updates, our oracle should be preferred w.r.t. the best-known (in a worst-case sense) MSF \emph{fully-dynamic} algorithm, requiring $O(\sqrt{n})$ time per update.
\end{abstract}

\section{Introduction}
Let $G=(V(G),E(G),w)$ be a positively real-weighted undirected graph of $n$ nodes and $m$
edges.
A \emph{shortest-path tree} (SPT) of $G$ rooted at a distinguished source vertex, say $s$, is one of the most
popular structures in communication networks. For example, it can be used for implementing the fundamental
\emph{broadcasting} operation. However, the SPT, as any tree-based topology, is highly sensitive to edge/vertex
failures, which cause the undesired effect of disconnecting sets of vertices from the source.

Therefore, a general approach to cope with this scenario is to make the SPT resistant against a given number of  component failures,
by adding to it a set of suitably selected edges from the underlying graph, so
that the resulting structure will still contain an SPT of the surviving network.
If we prepare ourselves to resist against a set of at most $f$
failing edges in $G$, then the corresponding structure will be named an \emph{f-edge-fault-tolerant} ($f$-EFT) SPT.
Unfortunately, it can be seen that even if $f = 1$ and $m = \Theta(n^2)$, then $\Theta(m)$ additional
edges may be needed, as will be shown in the full version of this paper.
Thus, to sparsify such a structure, it makes sense to resort to \emph{approximate} shortest paths from the source, that are stretched at most by a factor $\sigma>1$, for any
possible set of failures that has to be handled.

In this paper, we show how to build\footnote{Throughout this introduction, all the discussed structures are poly-time computable, even if we may omit to specify the actual running time.} an efficient structure of this sort. Moreover, we show that it is possible to transform such a structure into an efficient \emph{oracle} that will allow to quickly switch to the alternative paths in case a set of failures will take place.

\subsection{Related Work}
In the recent past, several single and
multiple edge/vertex-fault-tolerant \emph{approximate} SPT (ASPT) structures have been devised.
More formally, we say that a spanning
subgraph $H$ of $G$ is an $f$-EFT $\sigma$-ASPT  if it satisfies
the following condition: For each set of edges $F \subseteq E(G)$ of size at most $f$,
all the distances from the source $s$ in the subgraph $H-F=(V(G), E(H) \setminus F,w)$ are at most $\sigma$ times longer than the
corresponding distances in $G-F$.
Similar definitions can be given for the \emph{vertex-fault-tolerant} (VFT) case.

A natural counterpart of fault-tolerant SPT structures are fault-tolerant \emph{$\sigma$-stretched single-source distance oracles} ($\sigma$-SSDO in the following), i.e., \emph{compact}
data structures that can be built with a \emph{low} preprocessing
time, and that are able to \emph{quickly} return $\sigma$-approximate distances/paths from the source following a set of failures. Converting a fault-tolerant SPT into a corresponding SSDO with the very same stretch, and additionally having a small size and a fast query time, is a quite natural process, because of its practical usage: computing the alternative post-failure distances/paths on the structure may indeed be very time consuming. However, such a conversion process is not straightforward, in general, since it requires to exploit distance-related information that are instead implicit in the underlying structure, and this has to be done by optimizing the trade-off between the size and the query time of the oracle.

Turning back our attention to fault-tolerant SPT structures, their study originated in \cite{BK13}, where the authors built
in $O(m \log n + n \log^2 n)$ time a 1-VFT 3-SSDO of size $O(n \log n)$, and, for \emph{unweighted} graphs, in
$O(m\sqrt{n/\epsilon})$ time a 1-VFT $(1+\varepsilon)$-SSDO of size $O(\frac{n}{\varepsilon^3}+n \log n)$, for any
$\epsilon >0$, both having a distance (resp., path) query time of $O(1)$ (resp., proportional to the path's size). In
such a paper, the authors observe explicitly that, as a result of independent interest, the latter oracle (but actually
the former as well) can be converted into a corresponding structure (i.e., a spanning subgraph), having the same size and stretch.
For the weighted case, the obtained 1-VFT 3-ASPT of size $O(n \log n)$ was then substantially improved in \cite{BGLP14}, where the authors showed the existence of
a 1-E/VFT $(1+\varepsilon)$-ASPT of size $O(\frac{n \log n}{\varepsilon^2})$, for any $\varepsilon >0$ (without providing a corresponding oracle).

Concerning unweighted graphs, Parter and Peleg in \cite{PP13} presented a 1-E/VFT \emph{Breadth-First
Search tree} (BFS) of size $O(n \cdot \min\{\mathit{ecc}(s),\sqrt{n}\})$, where $\mathit{ecc}(s)$ denotes the eccentricity of the source vertex $s$ in $G$, namely a structure containing \emph{exact} shortest paths from the source after a single edge/vertex failure. In the same paper, the authors also exhibit a corresponding lower bound of $\Omega(n^{3/2})$ for the size of a 1-E/VFT BFS. Then, in \cite{PP14}, the same authors presented a
set of lower and upper bounds to the size of fault-tolerant $(\sigma,\beta)$-ABFS, where a further additive distortion $\beta$ is allowed to the distances.
More precisely, they showed that for every $\beta \in [1, O(\log n)]$, there exists a graph $G$ and a source vertex $s
\in V(G)$ such that a corresponding 1-EFT $(1,\beta)$-ABFS requires $\Omega(n^{1+\epsilon(\beta)})$ edges, for some
function $\epsilon(\beta) \in (0, 1)$. Moreover, they also constructed a 1-EFT $(1,4)$-ABFS of size $O(n^{4/3})$.
Finally, assuming at most $f=O(1)$ edge failures can take place, they showed the
existence of (i) an $f$-EFT $(3(f +1),(f+1) \log n)$-ABFS of size $O(fn)$, and (ii) an $f$-EFT $(3f+4)$-ABFS of size $O(fn \log^{f+1}n)$. These structures will be exactly our touchstone in this paper, since they are the only ones concerned with multiple-edge-failure single-source shortest paths.

\subsection{Our Results}
In this paper, we present the following main results:
\begin{itemize}
	\item An $f$-EFT $(2|F|+1)$-ASPT of size $O(f n)$ that is able to handle the failure of any set $F \subseteq E(G)$ of at most
		$f$ edges. This considerably improves w.r.t.\ to its direct
		competitors, namely the structures presented in \cite{PP14}: our structure has a size that is never worse, a lower
		stretch, works on weighted graphs, and handles an arbitrary (i.e., even non-constant)  number of failures.
		Moreover, our construction is simpler and can be computed quickly in $O(f m \, \alpha(m,n))$ time, where
		$\alpha$ is the inverse of the Ackermann's function.

	\item A corresponding $f$-EFT $(2|F|+1)$-SSDO of size $O(\min\{m,fn\} \log^2 n)$, that has a query time for a
		post-failure distance from the source of $O(|F|^2 \log^2 n)$, and is also able to report the corresponding path in
		the same time plus the path size. The preprocessing time is $O(f m\, \alpha(m,n)+fn \log^3 n)$. Notice that if one is
		willing to use $O(m \log^2 n)$ space, then our oracle will be prepared to handle any number of edge failures (i.e.,
		up to $m$).
\end{itemize}

Interestingly enough, the former result is obtained by posing a simple yet surprising relationship between the
structure of the replacement paths and the \emph{minimum spanning forest} (MSF) of an ad-hoc auxiliary graph. This
approach is also useful to develop the latter result, that is indeed obtained through an efficient updating of an
MSF after that a \emph{batch} of $k$ edge modifications (i.e., edge insertions, deletions and weight changes) are simultaneously performed. For this problem indeed we provide the following result:
\begin{itemize}
	\item a \emph{sensitivity oracle}\footnote{We use this noun for the oracle in accordance with its
		functionality of only reporting the updates in the MSF.} of size $O(m \log^2 n)$, that can be built in $O(m \log^3 n)$ time, and
		is able to report in $O(k^2 \log^2 n)$ time the (at most $2k$) edges either exiting from or entering into the MSF.
As a result of independent interest, it is worth noticing that our oracle can be used to efficiently maintain an MSF under relatively short sequences of \emph{non-simultaneous} updates. Indeed, observe that a sequence $\lambda=\langle \lambda_1, \ldots, \lambda_h
		\rangle$ of updates can be managed through $h$ sequential queries to the oracle, where the $i$-th query will involve
		the modifications to the starting MSF induced by the batch of the first $i$ updates. This
		way, we spend $O(h^2 \log^2 n)$ time to handle each single update.  Hence, as the fastest long-standing algorithm for the classic (and clearly more general) \emph{fully-dynamic} MSF problem has a worst-case cost of $O(\sqrt{n})$ per update \cite{EGIN92}, it follows that for $h=o(\sqrt[4]{n}/\log n)$, our oracle should be preferred, since it will manage each update in $o(\sqrt{n})$ time.
Notice also that a comparison with other known online/offline algorithms for maintaining an MSF that
		are more efficient in an amortized sense, like for instance those given in \cite{Eppstein94,HK01,HLT01}, is
		unfeasible, since they need to start from an empty graph to guarantee their bounds (or, they
		need long sequences of updates to become efficient). Thus, when starting from an arbitrary graph, as it
		happens in our setting, a single update operation could even cost them $\Theta(n)$ time!
\end{itemize}

Finally, we point out that we are also able to prove a lower bound of $\Omega(n^{1+\frac{1}{k}})$ on the size of any $f$-EFT $\sigma$-ASPT with $f \ge \log n$ and $\sigma < \frac{3k+1}{k+1}$, that holds if the Erd\H{o}s' girth conjecture is true. Our lower bound shows that, in contrast to the single-edge failure case, it is not possible to obtain a stretch arbitrary close to 1 with size $\softO(n)$ when the number of faults is more than $\log n$. We look at the problem of understanding if this can be done for constant $f>1$ as an interesting open problem. This result is given in Appendix \ref{sec:lower_bound}.

\subsection{Other Related Work on Fault-Tolerant Single/Multiple-Source Structures/Oracles}
Besides the papers mentioned before, several other research efforts have been devoted to structures and oracles for tolerating single/multiple failures in single-source shortest paths.
An early work on the topic is \cite{NPW03}, where the authors were concerned with the computation of \emph{best swap
edges} (w.r.t.\ several swap functions) for the failure of each and every edge in an SPT. As a by-product of their
results, it can be easily seen that by adding to an SPT the (at most) $n-1$ best swap edges w.r.t.\ to the new distance from $s$ to the root of the subtree disconnected from $s$ after an edge failure, then a 1-EFT 3-ASPT is obtained.
Interestingly, such a structure can be easily converted into a 1-EFT 3-SSDO of size $O(n)$ and query time $O(1)$.
Recently, in \cite{SIROCCO15}, the authors faced the special case of \emph{shortest-path failures}, in which the failure of a set $F$ of at most $f$ adjacent edges along any source-leaf path has to be tolerated. They proposed an $f$-EFT $(2k-1)(2|F|+1)$-ASPT of size $O(kn \,f^{1+1/k})$, where $|F|$ denotes the size of the actual failing path, and $k\geq 1$ is a parameter of choice. Notice that this result is subsumed by ours. Moreover, they also provided a conversion to a corresponding oracle, and for the special case of $f=2$, they gave an ad-hoc solution of size $O(n \log n)$ and with stretch 3.
For directed graphs with integer positive edge weights bounded by $M$, in \cite{GW12} the authors showed how to build efficiently in $\softO(M n^{\omega})$ time a randomized
1-EFT 1-SSDO of size $\Theta(n^2)$ and with $O(1)$ query time,
where returned distances are exact w.h.p., and $\omega< 2.373$ denotes the matrix
multiplication exponent.

Concerning unweighted graphs, in \cite{BGLP14} the authors showed that an \emph{ordinary} (i.e., non fault-tolerant)
$(\sigma,\beta)$-\emph{spanner} (i.e., where distances/paths between arbitrary pairs of nodes are at most
$(\sigma,\beta)$-stretched) of size $O(g(n))$ can be used to build a 1-EFT (resp., VFT) $(\sigma,\beta)$-ABFS of the
same size (resp., of size $O(g(n)+n \log n)$). This result is useful for building sparse 1-VFT $(1,\beta)$-ABFS
structures by making use of the vast literature on \emph{additive} $(1,\beta)$-spanners (e.g., \cite{BKMP10,Che13}).
Finally, Parter in \cite{Parter15} presented a 2-EFT BFS having $O(n^{5/3})$ edges, which is tight.

Another research stream related to our work is that on \emph{multi-source} (MS) fault-tolerant structures, for which we look at distances/paths from a set $S \subseteq V(G)$ of sources. Here, results are known only for unweighted graphs. In \cite{PP13} the authors gave an algorithm to compute a 1-EFT MSBFS of size $O(\sqrt{|S|} \, n^{3/2})$, which is tight.
Then, in \cite{BGLP14} it was shown that an ordinary $(\sigma,\beta)$-spanner of size $O(g(n))$ can be used to build a 1-EFT $(\sigma,\beta)$-AMSBFS of size $O(g(n)+n \,|S|)$, and similarly for the vertex case of size $O(g(n)+n \log n |S|)$.

\subsection{More Related Work on (Fault-Tolerant) Spanners/Oracles}
For the sake of completeness, we also give some hints on the large body of literature on the related topic of (fault-tolerant) spanners and distance oracles.

On weighted graphs, the currently best known construction is, for any $f \ge 1$ and any integer parameter $k \geq 1$, the $f$-EFT (resp., VFT) $(2k-1)$-spanner of size $O(f \, n^{1+1/k})$ (resp., $\softO(f^2 \, k^{f+1} \, n^{1+1/k})$) given in \cite{CLPR09}. For the vertex-failure case, this has been then improved in a randomized sense in \cite{DK11}, where the expected size was reduced to $\softO(f^{2-1/k} \, n^{1+1/k})$.
For a comparison, the sparsest known $(2k-1)$-multiplicative ordinary spanner has size $O(n^{1+1/k})$ \cite{DBLP:journals/dcg/AlthoferDDJS93}, and this is believed to be asymptotically tight due to the girth conjecture of Erd\H{o}s \cite{erdHos1964extremal}.
Then, in \cite{AFIR13} it was introduced the resembling concept of 1-EFT \emph{resilient} spanners, i.e., spanners such that whenever any edge in $G$ fails, then the relative distance increases in the spanner are very close to those in $G$.

Ordinary (i.e., fault-free) \emph{all-pairs distance oracles} (APDO) on weigthed graphs were introduced in a seminal work by Thorup and Zwick \cite{TZ05} (who also coined the term \emph{oracle}), followed by a sequel of papers (among the others, we mention \cite{DBLP:conf/stoc/Chechik14,EP15} for the currently best bounds).
In a fault-tolerant setting, in \cite{BK09} the authors built (on directed graphs)
a 1-E/VFT $1$-APDO of size $\softO(n^2)$ and with query time $O(1)$.
For two failures, in \cite{DP09} the authors built, still on directed graphs, a 2-E/VFT $1$-APDO of size $\softO(n^2)$ and with query time $O(\log n)$.
Concerning multiple-edge failures,  in \cite{CLPR10} the authors built, for any
integer $k \geq 1$, an $f$-EFT $(8k - 2)(f + 1)$-APDO of size $O(fk \, n^{1+1/k}
\log (n W))$, where $W$ is the ratio of the maximum to the minimum edge weight
in $G$, and with a query time of
$\softO(|F| \, \log \log d)$, where $F$ is the actual set of failing edges, and $d$ is the distance between the
queried pair of nodes in $G-F$.

On unweighted graphs, it makes instead sense to study fault-tolerant additive spanners. In particular, Braunshvig et al. \cite{BCP12} proposed the following general approach to build an $f$-EFT additive spanner: Let $A$ be an $f$-EFT $\sigma$-spanner, and let $B$ be an ordinary $(1,\beta)$-spanner. Then $H=A\cup B$ is an $f$-EFT $(1,2f(2\beta+\sigma-1)+\beta)$-spanner. Recently, in \cite{ESA15} the corresponding analysis has been refined yielding a better additive bound of $2f(\beta + \sigma - 1) + \beta$.
Finally, for other results on single edge/vertex failures spanners/oracles on unweighted graphs, we refer the reader to \cite{BK13,P14,ESA15}.

\section{An \texorpdfstring{$f$-EFT $(2|F|+1)$}{f-EFT (2|F|+1)}-ASPT and a Corresponding Oracle}
\label{sec:f_structure}

In this section we show how to compute an $f$-EFT $(2|F|+1)$-ASPT $H$ of $G$.
When up to $f$ edges can fail, it is easy to see that whenever $G$ is $(f+1)$-edge-connected, $H$ must contain
$\Omega(fn)$ edges even if we are only interested in preserving the connectivity of $G$, since the degree of each vertex must be at least equal to $f+1$.
Here we show that $|E(H)| = O(fn)$ edges also suffice if we seek to preserve distances that are at most $(2f+1)$-stretched w.r.t.\ the surviving
part of $G$.

Let $d_X(u,u')$ and $\pi_X(u, u^\prime)$ denote the distance and the shortest path between nodes $u$ and $u'$ in any
subgraph $X$ of $G$, respectively. When $u=s$, we will simply write $d_X(u^\prime)$ and $\pi_X(u^\prime)$.
If $\pi$ is a path, $\pi[u,u^\prime]$ will denote the subpath of $\pi$ between $u,u^\prime \in V(\pi)$.

For any given integer $f$,
Algorithm~\ref{alg:eft-spt} returns an $f$-EFT $(2|F|+1)$-ASPT of $G$.  First, it computes an SPT $T$ of $G$ that is
used to assign a weight to the edges of an auxiliary graph $G^\prime=(V(G), E(G),w')$. More precisely, the weight of an edge
$e$ of $G^\prime$ is $0$ if $e$ is also in $T$, otherwise it is equal to the sum of the corresponding edge weight in $G$ and the
distances in $T$ between $s$ and the endpoints of $e$. Then, $f+1$ MSFs $M_0, \dots, M_f$ of $G^\prime$ are iteratively
computed: when we compute the $i$-th forest, we remove its edges $M_i$ from $G^\prime$ before computing the $(i+1)$-th forest, so that the sets $M_i$ are pairwise disjoint.  The sought subgraph $H$ contains all the edges of the sets $M_i$.
Notice that $M_0$ coincides with $E(T)$.

\begin{algorithm}[t]
\footnotesize
\caption{Algorithm for computing an $f$-EFT $(2|F|+1)$-ASPT of $G$.}	
	\label{alg:eft-spt}

	\DontPrintSemicolon
	$T \gets$ compute an SPT of $G$ \;
	\For{$(u,v) \in E(G)$}
	{
		\lIf{$(u,v) \in E(T)$}
		{
			$w^\prime(u, v) \gets 0$
		}
		\lElse
		{
			$w^\prime(u,v) \gets d_T(u) + w(u,v)+ d_T(v)$
		}
	}
	$G^\prime \gets (V(G), E(G),w')$ \;
	\BlankLine
	
	$G_0 \gets G^\prime$ \;
	\For{$i = 0, \dots, f$}
	{
		$M_i \gets$ edges of an MSF of $G_i$ (w.r.t.\ $w^\prime$) \label{ln:compute_MST} \;
		$G_{i+1} \gets G_i \setminus M_i$ \;
	}

	\BlankLine
	
	$H \gets $ subgraph of $G$ containing the edges in $\bigcup_{i=0}^f M_i$ \;
	\Return $H$
\end{algorithm}

We now argue that $H$ is indeed an $f$-EFT $(2|F|+1)$-ASPT of $G$.
Fix a vertex $t$ and let $\pi = \pi_{G-F}(t)$ be the shortest path from $s$ to $t$ in the surviving graph $G-F$.\footnote{We assume that such a path exists, as otherwise $d_{G-F}(t)=+\infty$ which implies $d_{H-F}(t)=+\infty$, and we are done.}
The path $\pi$ traverses the vertices of several trees in the forest $T-F$.
We say that an edge is \emph{new} if its endpoints belong to two different trees in $T-F$.
Let $N$ be the set of new edges in $\pi$.

Now consider an MSF $M$ of the graph $H - F$ (w.r.t.\ $w^\prime$). This is also an MSF of the graph $G^\prime-F$
(w.r.t.\ $w^\prime$) as shown by the following lemma.
\begin{lemma}
	\label{lemma:M_mst_of_G-F}
	For every $F \subseteq E(G)$ with $|F| \le f$, any MSF $M$ of $H-F$ (w.r.t.\ $w^\prime$) is also an MSF of $G^\prime-F$ (w.r.t.\ $w^\prime$).
\end{lemma}
\begin{proof}
	In what follows, whenever ties arise we break them by prioritizing the edges in $H$.  First we
	show that, given any cut-set\footnote{A cut-set of a graph $X$ is a subset of $E(X)$ whose removal  increases the number of connected components of $X$.} $C$ of $G^\prime$, $H$ contains the $\min\{|C|, f+1\}$ lightest edges of $C$.  Indeed, for
	any set $M_i$, consider the set $C_i = C \setminus \cup_{j=0}^{i-1} M_j$. Either $C_i$ is non empty, and therefore
	$M_i$ contains the lightest edge in $C_i$, or $C_i = \emptyset$ which means that each edge in $C$ belongs to some
	set $M_j$ and hence to $H$.

	Let $M^\prime$ be an MSF of $G^\prime-F$. We prove the claim by showing that each edge $e \in E(M^\prime)$ must also
	belong to $M$.  Let $C^\prime$ be the cut-set of $G'$ that contains $e$ and every edge $e' \in  E(G')$ that forms a cycle with $e$ in $M' \cup \{ e^\prime \}$. Since $e$ is
	the lightest edge of $C^\prime \setminus F$, it is within the $f+1$ lightest edges of $C^\prime$. As a consequence
	$e \in E(H-F)$, and it also belongs to $M$ as it is the lightest edge in $C^\prime \cap E(H-F)$.
\end{proof}

Let $\pi^\prime = \pi_M(s, t)$ and notice that $\pi'$ traverses each tree of the forest $T-F$ at most once since edges in $E(T)$ have weight $0$ in $H$. Once again, let $N^\prime$ be the set of new edges of $\pi^\prime$.
We now provide an upper bound to the distance $d_{H-F}(t)$ using the path $\pi^\prime$:
\begin{lemma}
	\label{lemma:H-F_path_upperbound}
	$d_{H-F}(t) \le w(\pi^\prime) \le  \sum_{e \in N^\prime} w^\prime(e) + d_{G}(t)$.
\end{lemma}
\begin{proof}
	Let $M$ be an MSF of the graph $H-F$ (w.r.t. $w'$). The first inequality is trivial as $\pi^\prime=\pi_M(s,t)$ is a path (not necessarily shortest) between
	$s$ and $t$ in (a subgraph of) $H-F$, hence we focus on proving the second inequality.

	Let $T_0, \dots, T_h$ be the trees of $T-F$ traversed by $\pi^\prime$, in order, and let $e^\prime_i =
	(v_{i-1}, u_i)$ be the new edge in $\pi^\prime$ connecting a vertex $v_{i-1}$ of $T_{i-1}$ to a vertex $u_{i}$ of
	$T_i$. In such a way we have $N^\prime = \{e^\prime_1, \dots, e^\prime_h\}$.
	We call $r_i$ the vertex in $V(T_i) \cap V(\pi^\prime)$ that has the lowest depth in $T_i$.\footnote{We think of $T_i$
		as rooted in the vertex of $V(T_i)$ which is closest to $s$ in $T$.}
	According to this definition, $r_0$ coincides with $s$, $r_h$ is the lowest common ancestor between $u_h$ and $t$, and $r_i$ is the
	lowest common ancestor between $u_i$ and $v_i$, for every $0<i<h$.
	
	We prove by induction on $i$ that $w(\pi^\prime[s, r_i]) \le \sum_{j=1}^i w^\prime(e^\prime_j)$. The base case $i=0$	is trivially true.
	Now suppose that the inductive hypothesis holds for $i$, we prove it also for $i+1$:
	\begin{multline*}
		w(\pi^\prime[s, r_{i+1}]) = w(\pi^\prime[s, r_i]) + d_{T_i}(r_i, v_i) + w(e^\prime_{i+1}) +
		d_{T_{i+1}}(u_{i+1}, r_{i+1}) \\
		  \le \sum_{j=1}^{i} w^\prime(e^\prime_j) + d_{T}(v_i) + w(e^\prime_{i+1}) +	d_{T}(u_{i+1})  		
		 = \sum_{j=1}^{i} w^\prime(e^\prime_j) + w^\prime(e^\prime_{i+1}) = \sum_{j=1}^{i+1}
		w^\prime(e^\prime_j).
	\end{multline*}

	We now use the fact that $d_{T_h}(r_h, t) = d_{T}(r_h, t) = d_{G}(r_h, t)$ to prove the claim:
	\begin{align*}
		w(\pi^\prime) = w(\pi^\prime[s,r_h]) +  w(\pi^\prime[r_h,t]) \le \sum_{j=1}^{h} w^\prime(e^\prime_j) +
		d_{T_h}(r_h, t) \le \sum_{j=1}^{h} w^\prime(e^\prime_j) + d_G(t).  \tag*{\qedhere}
	\end{align*}
\end{proof}

Next lemma shows that the weights of the new edges of $\pi^\prime$ are, in turn, upper bounded by the weight of some new
edge of the path $\pi$.

\begin{figure}[t]
	\centering
	\includegraphics[scale=0.95]{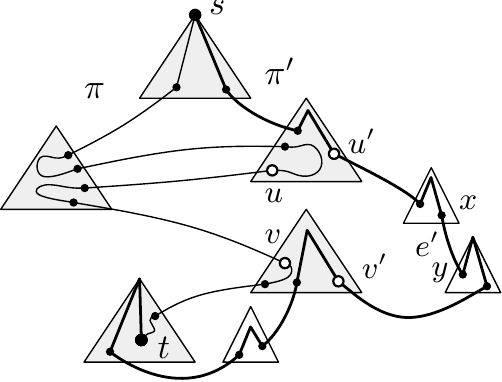}
	\caption{The forest $T-F$ obtained by deleting the failed edges in $F$ from $T$. The path $\pi$ is the shortest path
	between $s$ and $t$ in $G-F$, while $\pi^\prime$ (in bold) is the unique path in $M$ between the same vertices. Gray trees
	contain a vertex of $\pi$ and are therefore in $Z$. Edges having endpoints in different trees are \emph{new}.}
	\label{fig:spt_trees}
\end{figure}

\begin{lemma}
	\label{lemma:mst_edge_bound}
	For each $e^\prime \in N^\prime$, we have $w^\prime(e^\prime) \le \max_{e \in N} w^\prime(e)$.
\end{lemma}
\begin{proof}
	Let $e^\prime = (x, y)$ be an arbitrary edge in $N'$. W.l.o.g., we assume that the path $\pi'$ traverses the vertices $s$, $x$, $y$, $t$ in this order.
	We recall that the path $\pi^\prime$ traverses each tree in $T-F$ at most once, i.e., all the
	vertices of $\pi^\prime$ that belong to the same tree in $T-F$ must be contiguous in $\pi^\prime$. Moreover, as
	$e^\prime$ is new, $x$ and $y$ belong to two different trees in $T-F$.
	
Let $Z$ be the set of trees of the forest $T-F$ that are traversed by the path $\pi$. Let $u^\prime$ be the last vertex of $\pi^\prime[s,x]$ that belongs to a tree, say $T_u$, of $Z$ (see
	Figure~\ref{fig:spt_trees}). Observe that $u^\prime$ is always defined since $s$ belongs to some tree of $Z$. In a similar way, let
	$v^\prime$ be the first vertex of $\pi^\prime[y, t]$ that belongs to a tree, say $T_v$, of $Z$. Again, observe that $v^\prime$ is always defined as $t$ belongs to some tree of $Z$ other than that containing $s$, and so $T_u \neq T_v$, and finally notice that $e^\prime \in E(\pi'[u',v'])$.
	We know that $\pi$ traverses both $T_u$ and $T_v$ (in some order),  so we let $\pi^*$ be
	the minimal (w.r.t.\ inclusion) subpath of $\pi$ with one endpoint, say $u$, in $V(T_u)$, and the other endpoint, say $v$, in $V(T_v)$.

	Let $N^* = E(\pi^*) \cap N$ be the set of new edges in $\pi^*$. Notice that $N^* \neq \emptyset$ as $T_u \neq \T_v$, and that adding the edges in $N^*$  (weighted according to $w^\prime$) to $M$ forms (at least) a cycle $C$ containing both $e^\prime$ and an edge in $N^*$, say $e^*$.
	Since $M$ is an MSF of $G'-F$, as shown by Lemma~\ref{lemma:M_mst_of_G-F}, we have that $w^\prime(e^\prime) \le w^\prime(e^*) \le \max_{e \in N} w^\prime(e)$.
\end{proof}

Finally, next lemma relates the weights $w^\prime$ of the new edges of $\pi$ to distances in the surviving graph
$G-F$.
\begin{lemma}
	\label{lemma:sp_edge_bound}
	For $e \in N$, $w^\prime(e) \le 2d_{G-F}(t)$.
\end{lemma}
\begin{proof}
	Let $e=(u,v)$ with $d_{G-F}(u) \le d_{G-F}(v)$. Since $e$ lies on the shortest path $\pi=\pi_{G-F}(s,t)$, we can write:
	\[
		w^\prime(e) = d_T(u) + w(e) + d_T(v) \le d_{G-F}(v) + d_T(v) \le 2d_{G-F}(v) \le 2d_{G-F}(t). \tag*{\qedhere}
	\]
\end{proof}

We are now ready to prove the main result of this section:
\begin{theorem}
	The graph $H$ returned by Algorithm~\ref{alg:eft-spt} is an $f$-EFT $(2|F|+1)$-ASPT of $G$.
	Moreover, Algorithm~\ref{alg:eft-spt} requires $O(f m \,\alpha(m,n))$ time and $O(m)$ space.
\end{theorem}
\begin{proof}
	First, observe that $\pi^\prime = \pi_M(s,t)$ contains at most $|F|$ new edges. Indeed all the edges in $T-F$ have weight $0$, while the remaining
	edges have a positive weight. This means that $E(T-F) \subseteq E(M)$. As $T-F$ has no more than $|F|+1$ connected
	components, we have that at most $|F|$ other edges -- which are not in $E(T-F)$ -- can belong to $M$.
	
	By using the above fact in conjunction with Lemmas~\ref{lemma:H-F_path_upperbound}--\ref{lemma:sp_edge_bound}, we can write:
	\begin{align}
		\label{eq:spt_heaviest_mst_edge}
		\begin{split}
			d_{H-F}(t) 	& \le w(\pi^\prime) \le \sum_{e \in N^\prime} w^\prime(e) + d_{G}(t)
				\le |F| \max_{e \in N^\prime} w^\prime(e) + d_{G}(t) \\
				& \le |F| \max_{e \in N} w^\prime(e) + d_{G}(t)
				\le 2|F|d_{G-F}(t) + d_{G-F}(t) = (2|F| + 1) d_{G-F}(t).
		\end{split}
	\end{align}
	
	We recall that this holds for every vertex $t \in V(G)$.  Concerning the computational complexity of
	Algorithm~\ref{alg:eft-spt}, we make use of Chazelle's algorithm \cite{Cha00} -- that computes an MSF in $O(m\,\alpha(m,n))$ time and linear space -- to compute the $f+1$ MSFs $M_0,\dots,M_f$.
\end{proof}

\subsection{A Corresponding Oracle}
\label{sec:f_oracle}
In this section we show how to build an oracle that, given a positively real-weighted graph $G$ and a distinguished source vertex $s$, is able to answer queries of the form:
\emph{Given a set $F$ of at most $f$ edge failures, and a destination node $t$ in $G$, report a $(2|F|+1)$-approximate path/distance from $s$ to $t$ in $G-F$.}

We first compute an SPT $T$ of $G$ and a $f$-EFT $(2|F|+1)$-ASPT $H$ of $G$, as shown in the previous section. Then, the
oracle is composed of three ingredients:
\begin{itemize}
	\item the tree $T$ and all the distances $d_T(v) = d_G(v)$ from $s$ to any vertex $v \in V(G)$;
	\item an MSF sensitivity oracle $Q$ of $H$ w.r.t.\ the weights $w^\prime$, built as shown in Section~\ref{sec:MST_oracle};
	\item an oracle to answer \emph{lowest common ancestor} (LCA) queries between two vertices in $T$.
		Such an oracle can be built in linear time and has a constant query time \cite{HT84}.
\end{itemize}
The resulting size is $O(fn \log^2 n)$ and the time required to build our oracle is $O(fm\,\alpha(m,n) + fn \log^3 n)$.
Interestingly, if we do not know the value of $f$ in advance, we can build, in $O(m \log^3 n)$ time, an oracle of size $O(m \log^2 n)$ that is able to
report $(2|F|+1)$ approximate paths/distances, for \emph{any} number $|F|$ of faults.

We will make use of the following additional property of our MSF oracle $Q$, that will be shown in Section~\ref{sec:MST_oracle}:
\emph{$Q$ can report, in $O(|F|^2 \log^2 n)$ time, all the new
edges (and their weights), on the unique path from $s$ to $t$ in the updated MSF, in order.}

\paragraph*{Answering a Path Query.}
To return a $(2|F|+1)$-approximate path between $s$ and $t$, it suffices to
report the path $\pi^\prime = \pi_M(s,t)$, as shown by Equation~\eqref{eq:spt_heaviest_mst_edge}.

We query the MSF oracle $Q$ for the new edges on the unique path from $s$ to $t$ in the updated MSF.
Let $\langle e^\prime_1, \dots, e^\prime_h \rangle$ be these new edges, in order, with $e^\prime_i = (v_{i-1}, u_i)$.
For $0<i<h$, let $r_i$ be the LCA between $u_i$ and $v_i$, and let $r_h$ be the LCA between $u_h$ and $t$.
We now have all the pieces to reconstruct and return the path $\pi^\prime$.
Indeed, if we let $\pi^\prime_i = \pi_T(u_i, r_i) \circ \pi_T(r_i, v_i)$, the following holds:
\begin{equation}
	\label{eq:oracle_path_decomposition}
	\pi^\prime = \pi_T(s, v_0) \circ e^\prime_1 \circ \pi^\prime_1 \circ e^\prime_2 \circ \pi^\prime_2 \circ \dots \circ \pi^\prime_{h-1} \circ e^\prime_h \circ \pi_T(u_h, r_h) \circ \pi_T(r_h, t)
\end{equation}
where each subpath is entirely in $T$ and all the endpoints are known.
The whole procedure requires $O(|F|^2 \log^2 n)$ time to perform the query on $Q$, $O(|F|)$ time for the LCA queries,
and $O(|\pi^\prime|)$ time to reconstruct the path. The overall query time is therefore $O(|F|^2 \log^2n + |\pi^\prime|)$.

\paragraph*{Answering a Distance Query.}
To report the length of a $(2|F|+1)$-approximate path from $s$ to $t$, we can replace each subpath in Equation~\eqref{eq:oracle_path_decomposition} with the corresponding distance, in order to obtain:
\begin{equation*}
	w(\pi^\prime) = d_T(v_0) + \sum_{i=1}^{h-1} \bigg( w(e^\prime_{i}) + d_T(u_i, r_i) + d_T(r_i, v_i) + w(e^\prime_{i+1}) \bigg) + d_T(u_h, r_h) + d_T(r_h, t).
\end{equation*}
The above quantity can be computed in $O(h) = O(|F|)$ time, once we know the edges $e_1, \dots, e_h$
and we notice that $w(e^\prime_i) = w^\prime(e^\prime_i) - d_T(v_{i-1}) - d_T(u_i)$, and that if $x$ is a descendant of $r_i$ in $T$, then $d_T(r_i, x) = d_T(x)-d_T(r_i)$.
The overall query time is thus $O(|F|^2 \log^2 n)$.

\section{A Minimum Spanning Tree Sensitivity Oracle}
\label{sec:MST_oracle}
In this section we present an oracle that, given a real-weighted graph $G$ with $n$ vertices and $m$ edges, along with any \emph{minimum spanning tree} (MST) $T$ of $G$, is able to answer queries of the form:

\emph{``Given a set of  $k$ edge updates on $G$ (i.e., edge insertions, deletions and weight modifications), let
$T^\prime$ be the new MST of $G$. What are the edges in the symmetric difference of $E(T)$ and $E(T^\prime)$?''}\footnote{For the sake of avoiding technicalities, we assume that each edge is subject to at most a single update and we also assume that the graph $G$ always remains connected, so that we simply talk about an MST instead of an MSF of $G$. For instance, this can be easily guaranteed by adding a dummy vertex $x \not\in V(G)$ that is connected to all the vertices of $V(G)$ with edges of large weights.}

In other words, the oracle can report all the edges of $T$ that leave the MST as a consequence of the updates, along
with all the new edges in $T^\prime$ that enter the MST in their place.  The oracle requires $O(m \log^2 n)$ space and can be built in $O(m \log^2 n)$ space and $O(m \log^3 n)$ time, while a query involving $k$ updates can be answered in $O(k^2 \log^2 n)$ time and space.

	Our oracle exploits the fact that, when few updates are to be handled, the changes in the resulting MST will be small. This implies that large portions of $T$ and $T^\prime$ will coincide, and knowing these portions would allow us to save a considerable amount of work compared to the time needed to recompute $T^\prime$ from scratch. To this aim, we build a structure that maintains a set of connected subtrees of $T$ at different levels of granularity.
	
In details, we will use a hierarchical clustering of the vertices of $T$. Our clustering is inspired by the construction of \emph{topology trees} given in \cite{F85}. In \cite{F85}, the author solves the \emph{dynamic} MST problem by using a collection of topology trees that are built on top of an auxiliary graph representing shrunk components of $G$. We use our clustering in a different way and, as we do not need to support permanent updates of $G$, we are also able to simplify the construction. Due to space limitations, the full description of our construction will be given in the extended  version of the paper, while here we provide a sketch of it.

We start by describing the properties of our hierarchical clustering. Let $\Delta$ be the maximum degree\footnote{In order
to compute the clustering, the tree $T$ will be rooted. We still define the degree of a vertex $v$ in $T$ to be the number of edges that are incident $v$, including the edge from $v$ to its parent in $T$, if any.} of a vertex in $T$. Each cluster $C$ will have a level $\ell(C) \in \{0, \dots, L\}$, and we will call $\C_i$ the set of clusters of level $i$. Our clustering will guarantee that:
\begin{description}
	\item[P$_1$.] Clusters of each level $i$ are a partition of the vertices of $T$, i.e.,
	they are pairwise disjoint and $\bigcup_{C \in \C_i} C = V(G)$;
	\item[P$_2$.] The vertices in each cluster induce a connected component of $T$;
	\item[P$_3$.] Clusters of level $0$ are singletons, i.e., they contain a single vertex of $T$;
	\item[P$_4$.] There is only one cluster of level $L$ (and it coincides with $V(T)$);
	\item[P$_5$.] Each cluster of level $i \ge 1$ is the union of at least $2$ and at most $\Delta$ clusters of level $i-1$.
\end{description}
\begin{figure}[t]
	\centering
	\includegraphics[scale=.60]{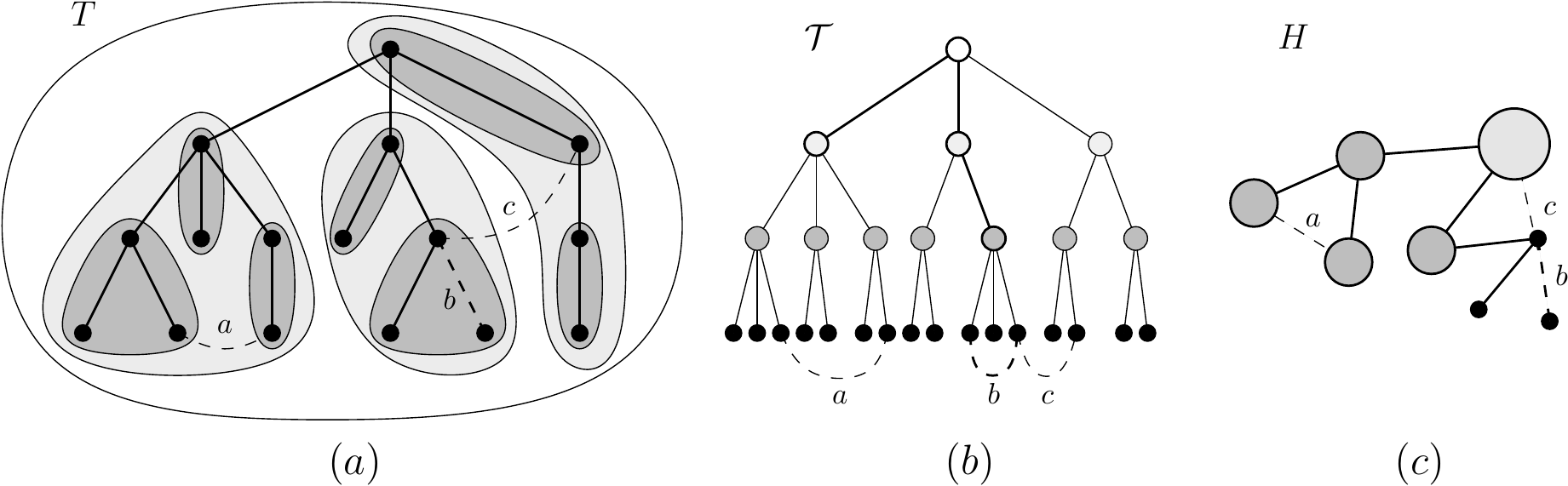}
	\caption{Hierarchical clustering of the vertices of $T$. (a) shows all the sets of $\C$, where black vertices are
	singletons. (b) shows the tree $\T$  associated with the clustering $\C$. (c) shows the graph $H$ whose vertices
	are computed by Algorithm~\ref{alg:split}. The edges in $F$ are depicted using dashed lines.}
	\label{fig:clustering}
\end{figure}
It follows from the above properties that a cluster of level $i$ contains at least $2^i$ vertices, and hence $L \le \log n$. Figure~\ref{fig:clustering}~(a) shows an example of such a clustering. This hierarchy can be represented by a tree $\T$ of height $L$ rooted in the unique cluster in $\C_L$. The children of a
cluster of level $i \ge 1$ in $\T$ are the clusters of level $i-1$ it contains (see Figure~\ref{fig:clustering}~(b)).

For each pair of clusters $C, C^\prime$ with $C \neq C^\prime$ we maintain an ordered set $E(C, C^\prime)$ containing all the edges of $E(G)$ with one endpoint in $C$ and the other in $C^\prime$. This set is ordered according to edge weights in a non-decreasing fashion.
Let $C(u)$ be the set of the $L+1$ clusters of the hierarchy (one for each level) that contain vertex $u$.
It is easy to see that an edge $(u,v) \in E(G)$ appears in at most $|C(u)| \cdot |C(v)| = O( \log^2 n )$ sets, and hence the overall number of elements in the sets is at most $O(m \log^2 n)$.

We now describe how a query can be answered. In order to do so, it is useful to split each weight update operation involving an edge $e$ into two separate operations, namely the deletion of $e$ followed by its reinsertion with the new (updated) weight. By doing so, all the operations in $F$ are now either insertions or deletions. For the sake of clarity, we first consider the case in which all the updates $F$ are edge deletions, and we will show later how this can be extended  to deal also with edge insertions. 

\paragraph*{Handling Edge Deletions.}
In order to handle deletions, we use Algorithm~\ref{alg:split} to construct an auxiliary graph $H$ whose vertices are clusters.
The algorithm will compute a set $R$ of clusters of $\T$ that will coincide with $V(H)$.
Initially $R$ contains the unique cluster in $C_L$ that is the root of $\T$ and represents the whole tree $T$. At each time, the set of clusters in $R$, although of different levels, will always form a partition of the vertices of $T$.
The algorithm proceeds iteratively, by considering one after the other the edges of $F$. When an edge $(u,v)$ is considered, if
$u$ and $v$ belong to the same cluster $C$ of $R$, we \emph{split} $C$, i.e., we remove $C$ from $\T$ and $R$, and we add
to $R$ all the clusters of level $\ell(C)-1$ contained in $C$. In this way $\T$ is always a forest and $R$ contains the
roots of its trees.

\begin{algorithm}[t]
\footnotesize
\caption{Algorithm for computing the set of vertices (i.e., clusters) of $H$.}
	\label{alg:split}

	\DontPrintSemicolon
	
	$R \gets \C_L$ \;
	\For{$(u,v) \in F$}
	{
		\While{$\troot_\T(u) = \troot_\T(v)$}
		{
			$C \gets \troot_\T(u) $ \;
			$\T \gets \T \setminus \{ C \}$ \;
			$R \gets (  R \setminus \{ C \} ) \cup \children_\T( C ) $ \tcp*{Split $C$}
		}
	}
	\Return $R$ \;
\end{algorithm}

In the end, $H$ is such that all the edges in $F$ have their endpoints into different clusters of $V(H)$. Moreover, as
each edge in $F$ can produce at most $L$ splits, and each split operation can increase the number of vertices by at most
$\Delta-1$, we have that $H$ contains at most $f \, L \, (\Delta-1) = O( \Delta \, f \log n)$ vertices (see
Figure~\ref{fig:clustering}~(c)).

To construct the set $E(H)$ we consider all the pairs $C,C^\prime$ of vertices in $V(H)$. For each of these pairs we
examine the edges in $E(C,C^\prime)$, in order, and we select the first edge $e$ so that $e \not\in F$, if any. Then, if $e$ exists, we add the edge $(C,C^\prime)$ to $H$ with weight $w(e)$.

We can now compute an MST $\wt{T}$ of $H$ in time $O( \Delta^2 \, f^2 \log^2 n )$ by using any standard MST algorithm.
Finally, we look at the edges of $\wt{T}$ and we answer the query by returning the edges in $E(\wt{T})$ that are not in $E(T)$.  Notice also that, once $\wt{T}$ has been computed, it is easy to report all the edges in $E(\wt{T}) \setminus
E(T)$ that belong to the unique path between any two vertices in the updated MST. This kind of query can still be
answered in $O( \Delta^2 \, f^2 \log^2 n )$ time, and it is needed by our fault-tolerant ASPT oracle of
Section~\ref{sec:f_oracle}.

\paragraph*{Handling General Edge Updates.}
It turns out that the complexity of the problem lies in handling the edge-deletion operations. Indeed, once this has been done, the remaining edge-insertion operations can be easily performed. To this aim, we reorganize the batch by first performing all the delete operations, and we make use
of a \emph{top-tree} \cite{AHLT05}, i.e., a data structure that dynamically maintains a (weighted) forest under edge-insertion (\emph{link}) and edge-deletion (\emph{cut}) operations. Moreover, given two vertices $u$ and $v$, top-trees are able to report the heaviest edge that lies on the path between $u$ and $v$ in the current forest. Each of these operations can be performed in $O(\log \eta)$ time where $\eta$ is the number of vertices of the forest.

The idea is to maintain the current MST $T^\prime$ by using a top-tree that is initialized when the oracle is built to represent the tree $T$. This takes $O(n \log n)$ time.
Then, we perform all the edge-deletion operations (as already described), while updating the top-tree accordingly (this requires $O(|F| \log n)$ time since the number of needed link and cut operations is $O(|F|)$).

Now we handle the insertions one by one. In order to insert a new edge $e=(u,v)$, we  search for the heaviest edge $e^\prime$ of the path connecting $u$ and $v$ in $T^\prime$. If $e^\prime$ is heavier than $e$, we cut $e^\prime$ from $T^\prime$ and we link the two resulting components by adding the edge $e$. It is easy to see that this procedure requires an overall time of $O(|F| \log n)$.

By keeping track of all the $O(|F|)$ updates in the MST $T^\prime$, we can easily answer a query consisting of both edge-insertion and edge-deletion operations in $O( \Delta^2 \, f^2 \log^2 n )$ time.

\paragraph*{Reducing the Degree of $T$.}
So far, the complexity of our oracle depends on the maximum degree $\Delta$ of a vertex in $T$.
However, using standard techniques (see, e.g., \cite{F85}), we now show that the updates on the original graph $G$ and its MST $T$
can be mapped onto an auxiliary graph $\bar{G}$ with weight function $\hat{w}$ and a corresponding MST $\hat{T}$, such that $\hat{G}$ has asymptotically the same size of $G$, and each vertex of $\hat{G}$ has a degree at most $3$ in $\hat{T}$.

\begin{figure}[b]
	\centering
	\includegraphics[scale=0.83]{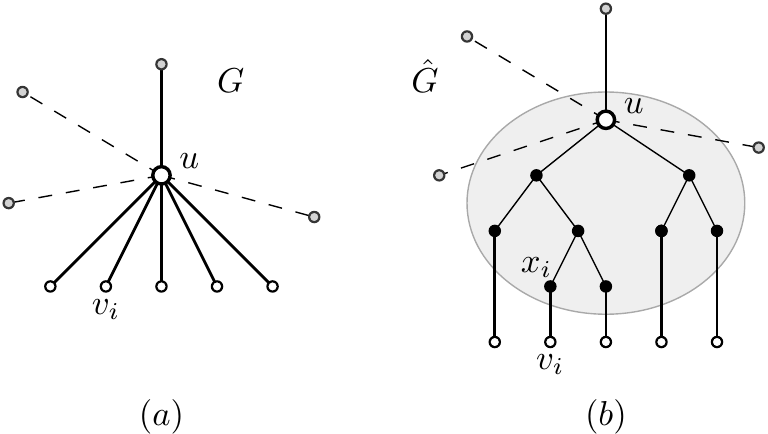}
	\caption{Reducing the degree of vertices in $T$: on the left side, the tree $T$ (solid edges) embedded in $G$, on
	the right side the superimposition of the binary tree to $T$ in order to get a maximum degree of 3. Thin solid edges have weight $0$, while the
weight of $(x_i, v_i)$ is $w(u. v_i)$.}
	\label{fig:degree}
\end{figure}

Initially $\hat{G}$, $\hat{w}$, and $\hat{T}$ coincide with $G$, $w$, and $T$, respectively. We iteratively search for a vertex $u$ in $\hat{T}$ that has more than $2$ children, and we lower its degree. Let $\children_{\hat{T}}(v) = \{ v_1, \dots, v_h \}$, we proceed as follows:
we remove all the edges in $\{ (u, v_i) : 1 \le i \le h \}$ from both $\hat{G}$ and $\hat{T}$, then we add to both $\hat{G}$ and $\hat{T}$ a binary tree whose root coincides with $u$, and that has exactly $h$ leaves $x_1, \dots, x_h$. We assign weight $\hat{w}(e)=0$ to all the edges $e$ of this tree.
Finally, we add to $\hat{G}$ and $\hat{T}$ an edge $(x_i, v_i)$ for each $1 \le i\le h$, and we set $\hat{w}(x_i, v_i) = w( u, v_i )$. An example of such a transformation is shown in Figure~\ref{fig:degree}.

Each time we have to perform a weight update or delete operation on an edge $(u,v_i)$ of $G$, we instead perform it on the corresponding edge $(x_i, v_i)$. Insertions and operations involving edges in $E(G) \setminus E(T)$ do not require any special care. In a similar way, whenever the answer of a query contains an edge $(x_i, v_i)$, we replace it with the corresponding edge $(u,v_i)$. Clearly, $O(n)$ vertices and edges are added by this process, and hence $|V(G)|=\Theta(|V(\hat{G})|)$ and $|E(G)|=\Theta(|E(\hat{G})|)$.

Once the maximum degree of the tree has been reduced to a constant, the query time
of our oracle becomes $O(f^2 \log^2 n)$. To achieve such a query time, however, we must
be careful in our implementation as it is discussed in Appendix~\ref{sec:implementation}.

\clearpage
\bibliographystyle{plain}
\bibliography{biblio}

\begin{thebibliography}{10}

\bibitem{AHLT05}
Stephen Alstrup, Jacob Holm, Kristian de~Lichtenberg, and Mikkel Thorup.
\newblock Maintaining information in fully dynamic trees with top trees.
\newblock {\em {ACM} Transactions on Algorithms}, 1(2):243--264, 2005.

\bibitem{DBLP:journals/dcg/AlthoferDDJS93}
Ingo Alth{\"{o}}fer, Gautam Das, David~P. Dobkin, Deborah Joseph, and
  Jos{\'{e}} Soares.
\newblock On sparse spanners of weighted graphs.
\newblock {\em Discrete {\&} Computational Geometry}, 9:81--100, 1993.

\bibitem{AFIR13}
Giorgio Ausiello, Paolo~Giulio Franciosa, Giuseppe~Francesco Italiano, and
  Andrea Ribichini.
\newblock On resilient graph spanners.
\newblock In {\em ESA}, pages 85--96, 2013.

\bibitem{BK13}
Surender Baswana and Neelesh Khanna.
\newblock Approximate shortest paths avoiding a failed vertex: Near optimal
  data structures for undirected unweighted graphs.
\newblock {\em Algorithmica}, 66(1):18--50, 2013.

\bibitem{BKMP10}
Surender Baswana, Kavitha Telikepalli, Kurt Mehlhorn, and Seth Pettie.
\newblock Additive spanners and (alpha, beta)-spanners.
\newblock {\em {ACM} Transactions on Algorithms}, 7(1):5, 2010.

\bibitem{BK09}
Aaron Bernstein and David~R. Karger.
\newblock A nearly optimal oracle for avoiding failed vertices and edges.
\newblock In {\em STOC}, pages 101--110, 2009.

\bibitem{ESA15}
Davide Bil{\`{o}}, Fabrizio Grandoni, Luciano Gual{\`{a}}, Stefano Leucci, and
  Guido Proietti.
\newblock Improved purely additive fault-tolerant spanners.
\newblock In {\em ESA}, pages 167--178, 2015.

\bibitem{BGLP14}
Davide Bil{\`{o}}, Luciano Gual{\`{a}}, Stefano Leucci, and Guido Proietti.
\newblock Fault-tolerant approximate shortest-path trees.
\newblock In {\em ESA}, pages 137--148, 2014.

\bibitem{BCP12}
Gilad Braunschvig, Shiri Chechik, and David Peleg.
\newblock Fault tolerant additive spanners.
\newblock In {\em WG}, pages 206--214, 2012.

\bibitem{Cha00}
Bernard Chazelle.
\newblock A minimum spanning tree algorithm with inverse-ackermann type
  complexity.
\newblock {\em J. {ACM}}, 47(6):1028--1047, 2000.

\bibitem{Che13}
Shiri Chechik.
\newblock New additive spanners.
\newblock In {\em SODA}, pages 498--512, 2013.

\bibitem{DBLP:conf/stoc/Chechik14}
Shiri Chechik.
\newblock Approximate distance oracles with constant query time.
\newblock In {\em STOC}, pages 654--663, 2014.

\bibitem{CLPR09}
Shiri Chechik, Michael Langberg, David Peleg, and Liam Roditty.
\newblock Fault-tolerant spanners for general graphs.
\newblock In {\em STOC}, pages 435--444, 2009.

\bibitem{CLPR10}
Shiri Chechik, Michael Langberg, David Peleg, and Liam Roditty.
\newblock \emph{f}-sensitivity distance oracles and routing schemes.
\newblock In {\em ESA}, pages 84--96, 2010.

\bibitem{SIROCCO15}
Annalisa D'Andrea, Mattia D'Emidio, Daniele Frigioni, Stefano Leucci, and Guido
  Proietti.
\newblock Path-fault-tolerant approximate shortest-path trees.
\newblock In {\em {SIROCCO}}, pages 224--238, 2015.

\bibitem{DK11}
Michael Dinitz and Robert Krauthgamer.
\newblock Fault-tolerant spanners: better and simpler.
\newblock In {\em PODC}, pages 169--178, 2011.

\bibitem{DP09}
Ran Duan and Seth Pettie.
\newblock Dual-failure distance and connectivity oracles.
\newblock In {\em SODA}, pages 506--515, 2009.

\bibitem{EP15}
Michael Elkin and Seth Pettie.
\newblock A linear-size logarithmic stretch path-reporting distance oracle for
  general graphs.
\newblock In {\em SODA}, pages 805--821, 2015.

\bibitem{Eppstein94}
David Eppstein.
\newblock Offline algorithms for dynamic minimum spanning tree problems.
\newblock {\em J. Algorithms}, 17(2):237--250, 1994.

\bibitem{EGIN92}
David Eppstein, Zvi Galil, Giuseppe~F. Italiano, and Amnon Nissenzweig.
\newblock Sparsification-a technique for speeding up dynamic graph algorithms
  (extended abstract).
\newblock In {\em FOCS}, pages 60--69, 1992.

\bibitem{erdHos1964extremal}
Paul Erd{\H{o}}s.
\newblock Extremal problems in graph theory.
\newblock In {\em Theory of Graphs and its Applications}, pages 29--36, 1964.

\bibitem{F85}
Greg~N. Frederickson.
\newblock Data structures for on-line updating of minimum spanning trees, with
  applications.
\newblock {\em {SIAM} J. Comput.}, 14(4):781--798, 1985.

\bibitem{GW12}
Fabrizio Grandoni and Virginia~Vassilevska Williams.
\newblock Improved distance sensitivity oracles via fast single-source
  replacement paths.
\newblock In {\em FOCS}, pages 748--757, 2012.

\bibitem{HMP01}
Torben Hagerup, Peter~Bro Miltersen, and Rasmus Pagh.
\newblock Deterministic dictionaries.
\newblock {\em J. Algorithms}, 41(1):69--85, 2001.

\bibitem{HT84}
Dov Harel and Robert~Endre Tarjan.
\newblock Fast algorithms for finding nearest common ancestors.
\newblock {\em {SIAM} J. Comput.}, 13(2):338--355, 1984.

\bibitem{HK01}
Monika~Rauch Henzinger and Valerie King.
\newblock Maintaining minimum spanning forests in dynamic graphs.
\newblock {\em {SIAM} J. Comput.}, 31(2):364--374, 2001.

\bibitem{HLT01}
Jacob Holm, Kristian de~Lichtenberg, and Mikkel Thorup.
\newblock Poly-logarithmic deterministic fully-dynamic algorithms for
  connectivity, minimum spanning tree, 2-edge, and biconnectivity.
\newblock {\em J. {ACM}}, 48(4):723--760, 2001.

\bibitem{NPW03}
Enrico Nardelli, Guido Proietti, and Peter Widmayer.
\newblock Swapping a failing edge of a single source shortest paths tree is
  good and fast.
\newblock {\em Algorithmica}, 35(1):56--74, 2003.

\bibitem{P14}
Merav Parter.
\newblock Vertex fault tolerant additive spanners.
\newblock In {\em DISC}, pages 167--181, 2014.

\bibitem{Parter15}
Merav Parter.
\newblock Dual failure resilient {BFS} structure.
\newblock In {\em PODC}, pages 481--490, 2015.

\bibitem{PP13}
Merav Parter and David Peleg.
\newblock Sparse fault-tolerant {BFS} trees.
\newblock In {\em ESA}, pages 779--790, 2013.

\bibitem{PP14}
Merav Parter and David Peleg.
\newblock Fault tolerant approximate {BFS} structures.
\newblock In {\em SODA}, pages 1073--1092, 2014.

\bibitem{TZ05}
Mikkel Thorup and Uri Zwick.
\newblock Approximate distance oracles.
\newblock {\em J. {ACM}}, 52(1):1--24, 2005.

\end{thebibliography}

\clearpage
\appendix

\section{Computing a Hierarchical Clustering of \texorpdfstring{$T$}{T}}
\label{sec:computing_clustering}
We now show how a clustering of $T$ satisfying properties P$_1$--P$_5$ of Section~\ref{sec:MST_oracle} can be found.  Starting from $C_0$, we build the
clusters of the hierarchy $\C$ in order of level, i.e., all the clusters in the set $\C_i$ are computed before the
clusters in the set $\C_{i+1}$.  At each time we maintain a rooted tree $T_i$ whose vertices represent clusters of level
$i$.  Initially $T_0 = T$ as each cluster in $C_0$ is a singleton, and $T_0$ is rooted in an arbitrary vertex $r_0$. At
each phase $i$, starting from $i=1$, we compute the clustering $\C_i$ by suitably partitioning the vertices of
$T_{i-1}$, as explained below. Then, all the vertices of $T_{i-1}$ that belong to the same cluster $C \in \C_i$ are identified into a
single vertex representing the whole cluster $C$. The resulting graph will be a tree $T_i$, that is now rooted
in the cluster $r_i$ containing $r_{i-1}$. This procedure is repeated until $T_i$ contains only a single vertex,
which represents $\C_L$.

It remains to describe how the partition $\C_i$ of $V(T_i)$ is computed: we start with $\C_i = \emptyset$ and
iteratively search for an internal vertex $v$ of $T_i$ of maximum depth. If $T_i$ only contains $v$, $v$'s children, and
$v$'s parent, we select $C=V(T)$ to be a new cluster of level $i$, otherwise we let $C= \{ v \} \cup \children_{T_i}(v)$.
We now add $C$ to $\C_i$, we remove the vertices of $C$ from $T_i$, and we search for a new cluster $C$ to add. We stop
as soon as $T_i$ becomes empty.

It is easy to see that all the properties P$_1$-P$_5$ are satisfied by this construction. Moreover, computing the set
$\C_i$ (and the tree $T_{i+1})$ from the tree $T_i$ takes linear time in $|V(T_i)|$. Since each vertex in $T_{i+1}$
represents at least two vertices in $T_i$, we know that $|V(T_{i+1})| \le \frac{1}{2} |V(T_i)|$, and we can easily
conclude that computing the whole hierarchical clustering takes linear time in $|V(T)|$.

\section{Implementation Details}
\label{sec:implementation}
As said in the paper, once the maximum degree of the tree has been reduced to a constant, the query time
of our oracle becomes $O(f^2 \log^2 n)$.  In order to achieve such a query time, however, we need to access in constant
time the ordered list of edges that cross the cut between any pair of clusters of $\C$.  Notice that explicitly storing
such a list for every pair of clusters would require $\Omega(n^2)$ space, in contrast with the claimed $O(m \log^2 n)$ space,
as empty lists contribute to space occupancy as well.

In order for our construction to provide the stated time and space bounds, we need to be careful in building our oracle.
Here we discuss how this can be done.  During the construction we will maintain a dictionary $D$, whose keys will be
pair of clusters and whose values will be pointers to the corresponding lists of edges. Initially $D$ is empty.  We sort
all the edges of $G$ in non-decreasing order of weight and we examine one edge at a time. When $e=(u,v)$ is considered, we
use the tree $\T$ to find all the clusters $C^1_u, C^2_u, \dots$ and  $C^1_v, C^2_v, \dots$ to whom $u$ and $v$ belong,
respectively. We stop just before reaching the LCA of $u$ and $v$ in $\T$ so that no cluster $C^i_u$ or $C^i_v$ contains
both $u$ and $v$. For each pair $\langle C^i_u, C^j_v \rangle$ we query $D$: if the key $\langle C^i_u, C^j_v \rangle$
exists, then we add $e$ to the corresponding list, otherwise we create a new list $L$ containing $e$ and we add to $D$ a
new element with $\langle C^i_u, C^j_v \rangle$ as its key and a pointer to $L$ as its value.

The above procedure requires $O(m \log^3 n)$ time, as each vertex belongs to $O(\log n)$ clusters and a query on $D$ requires $O(\log n)$ time. In order to reduce the query time, we now build a \emph{static} version of the dictionary $D$ that has \emph{constant} query time and linear size. This can be done in $O(\eta \log \eta)$ time where $\eta$ is the number of elements \cite{HMP01}. In our case $\eta=O(m \log^2 n)$, hence the overall building time becomes $O(m \log^3 n)$, while the size of the resulting structure is $O(m \log^2 n)$.

\section{A Lower Bound to the Size of a \texorpdfstring{$(\log n)$-EFT $\sigma$-ASPT}{(log n)-EFT sigma-ASPT} }
\label{sec:lower_bound}
In this section we show that, if the long-standing girth conjecture of Erd\H{o}s \cite{erdHos1964extremal} is true, then any $f$-EFT $\sigma$-ASPT with $f \ge \log n$ and $\sigma < \frac{3k+1}{k+1}$, requires $\Omega(n^{1+\frac{1}{k}})$ edges. In particular, this implies that if we want to be resistant to at least $\log n$ edge failures and to have stretch less than 2, then $\Theta(n^2)$ edges are needed.

\begin{figure}[ht]
	\centering
	\includegraphics[scale=1]{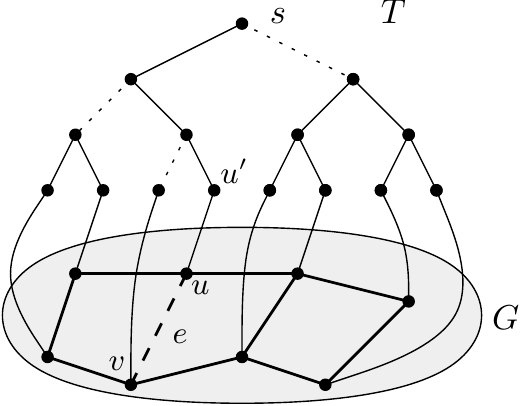}
	\caption{Graph $G^\prime$ used in the lower bound construction. The dashed edge $e$
	does not belong to $H$ while the dotted edges belong to $F$. Bold edges have weight $1$, tree edges have weight $0$,
	and the remaining edges --connecting the leaves of $T$ to the vertices in $G$-- have weight $x$.}
	\label{fig:lower_bound}
\end{figure}

Let $G$ be a graph on $\eta$ vertices with girth $g=2k+2$ and $\Omega(\eta^{1+\frac{1}{k}})$ edges (according to the girth conjecture, such a graph always exists).
We construct a weighted graph $G^\prime$ in the following way (see Figure~\ref{fig:lower_bound}): we add to $G$ a binary
tree $T$ rooted in $s$ with $\eta$ leaves and height $h= \left\lceil \log \eta \right\rceil$, and we further add an edge from each leaf of $T$ to a distinct vertex of $V(G)$, in an arbitrary way.
The weights of $E(G)$ and $E(T)$ will be set to 1 and 0, respectively, while the remaining additional edges will have weight $x = \frac{g}{2}-1$.
Observe that the total number of vertices of $G^\prime$ is $n=3\eta - 1$, hence $|E(G')|=\Omega(n^{1+\frac{1}{k}})$.

Let $H$ be any $f$-EFT $\sigma$-ASPT of $G^\prime$ rooted in $s$, with $f \ge \log n$ and $\sigma < \frac{3k+1}{k+1}$. We will show that $H$ must contain all the edges of $E(G)$.
Indeed, suppose that an edge $e=(u,v) \in E(G)$ is missing from $H$, and let $u^\prime$ be the unique leaf of the $T$ such that $(u, u^\prime) \in E(G^\prime)$. We let $\langle s=u_0, u_1, \dots, u_k \rangle$ be the sequence of internal vertices of $T$ traversed by $\pi_T(s, u^\prime)$, and let $e_i$ be the edge incident to $u_i$ other than that in $E(\pi_T(s, u^\prime))$. We choose $F = \{ e_0, e_1, \dots, e_k \}$ as shown in Figure~\ref{fig:lower_bound}.
It is easy to see that $|F| \le h = \left\lceil \log \eta  \right\rceil \le \log \eta +1 \le \log n$, and that each path from $s$ to any vertex of $V(G)$ in $G^\prime - F$ has the path $\pi = \pi_T(s,u^\prime) \circ (u^\prime, u)$ as a prefix, hence the same must hold in $H-F$.
Therefore, we know that $\pi_{H-F}(s, v) = \pi \circ \pi_{H-F}(u, v)$.
Observe that either $\pi_{H-F}(u, v)$ passes through a vertex in $V(T)$ or not.
In the former case, it must contain at least an edge of weight $1$ and two edges of weight $x$, hence $w(\pi_{H-F}(u, v)) \ge 2x+1 = g-1$. Otherwise, since the girth of $G$ is $g$, $w(\pi_{H-F}(u, v)) \ge g-1$.
In both cases we have that $d_{H-F}(v) = w(\pi_{H-F}(s, v)) = w(\pi) + w(\pi_{H-F}(u, v)) \ge \frac{g}{2} -1 +  g - 1 = \frac{3}{2}g -2$. At the same time, it holds $d_{G^\prime-F}(v) = w(\pi \circ (u,v)) = \frac{g}{2}$.
This implies that the stretch factor of $H$ would be at least $3 - \frac{4}{g} = \frac{3k+1}{k+1}$, a contradiction.

\end{document}